\documentclass[aps,prl,twocolumn,showpacs,preprintnumbers,amsmath,amssymb,divps]{revtex4-1}
\usepackage{graphicx}
\usepackage{dcolumn}
\usepackage{bm}

\usepackage{multirow}
\usepackage{epsfig}
\usepackage{amssymb}

\begin{document}

\title{Absence of Dirac Electrons in Silicene on Ag (111) Surfaces}

\author{Zhi-Xin Guo}
\email[On leave from  Xiangtan University, Xiangtan, Hunan 411105, China]{}
\author{Shinnosuke Furuya}
\author{Jun-ichi Iwata}
\author{Atsushi Oshiyama}

\affiliation{Department of Applied Physics, The University of Tokyo, Tokyo 113-8656, Japan}


\date{\today}

\begin{abstract}

We report first-principles calculations that clarify stability and electronic structures of silicene on Ag(111) surfaces. We find that several stable structures exist for silicene/Ag(111), exhibiting a variety of images of scanning tunneling microscopy. We also find that Dirac electrons are {\em absent} near Fermi energy in all the stable structures due to buckling of the Si monolayer and mixing between Si and Ag orbitals. We instead propose that either BN substrate or hydrogen processing of Si surface is a good candidate to preserve Dirac electrons in silicene.
\end{abstract}


\maketitle


Graphene is a hexagonally bonded carbon sheet in which $sp^2$ hybridized electrons ($\sigma$ electrons) form a honeycomb structure and the remaining $\pi$ electrons follow the massless Dirac (Weyl) equation. The energy bands show the linear dispersion (Dirac cone) at the Fermi level ($E_{\rm F}$) at particular symmetry points, $K$ and $K^{\prime}$, in Brilloine zone (BZ) \cite{SW}. This causes fascinating properties such as the anomalous quantum Hall effect \cite{geim,kim} and the unexpected magnetic ordering \cite{louie,okada}. Moreover, the in-plane high mobility intrinsic to graphite and its robustness place graphene as an emerging material to develop post-scaling semiconductor technology challenged by the cutting edge miniaturization \cite{itrs}.

In order to maximize technology connectivity, however, a hexagonal sheet consisting of Si which is located just below C in the periodic table has obvious superiority. Spin-orbit interaction stronger in Si than in C, possibly inducing the quantum spin-Hall phase \cite{kane}, is another intriguing factor in science viewpoints. Several experimental efforts to synthesize such a Si sheet (silicene) has been indeed done recently \cite{padova1,padova2}, and buckled silicene on Ag \cite{vogt,lin,jamagotchian} and on ZrB$_2$ \cite{fleurence} surfaces are identified by scanning tunneling microscopy (STM) measurements.

Calculations \cite{cahangirov} in the local density approximation (LDA) in the density-functional theory (DFT) have clarified that a hexagonally bonded flat Si sheet is unstable to the buckling and the resultant, still hypothetical, free-standing silicene with the buckling of 0.44 {\AA} shows the Dirac cone as in flat graphene. It is therefore extremely important to reveal whether Dirac electrons exist in the real silicene on the substrates.

Experimentally, the Si layer on the Ag(111) surface has been reported to exhibit several structures with the super-periodicity of 4$\times$4 \cite{vogt,lin,jamagotchian}, $\sqrt{13}\times\sqrt{13}$ \cite{lin,jamagotchian},  $2\sqrt{3}\times2\sqrt{3}$ \cite{jamagotchian,lalmi,feng} and $\sqrt{3}\times\sqrt{3}$ \cite{chen}  with respect to $1\times1$ of the Ag(111) surface. The former two have been well identified, whereas the latter two are still of controversy in terms of lattice mismatching \cite{vogt,lin}. Angular-resolved photoelectron spectroscopy (ARPES) measurements \cite{vogt} suggest the existence of the linear-dispersion band near $E_{\rm F}$ but the spectra are not conclusive.

In this Letter, we report extensive calculations for silicene on Ag(111) surfaces with the periodicity of 4$\times$4 and $\sqrt{13}\times\sqrt{13}$ as well as $2\sqrt{3}\times2\sqrt{3}$. We find six distinctive stable structures with the total-energy difference ranging in 69 meV per Si atom. Calculated scanning tunneling microscopy (STM) images of the most stable structures are in excellent agreement with the experiments available. An important finding common to all the (meta)stable structures is the interaction between silicene and Ag substrate, leading to the {\em disappearance} of the Dirac electrons near $E_F$. We instead propose that the hexagonal BN (h-BN) substrate and the hydrogen-processed Si \cite{smart-cut} are promising to preserve Dirac electrons in silicene.

Calculations are performed by our real space DFT (RSDFT) code \cite{IwataJCP}. The LDA is used to the exchange-correlation energy, and the norm-conserving pseudopotentials are adopted \cite{TM} to describe the electron-ion interaction. The grid spacing in the real-space calculations is taken to be 0.16 {\AA} corresponding to a cutoff energy of 108 Ry in the plane-wave-basis-set calculations. The silicene-covered Ag surface is simulated by a repeating slab model in which five-atomic-layer slabs are separated from each other by a 14-{\AA} vacuum region. The geometry optimization is performed until the remaining forces become less than 26 meV/{\AA} \cite{foot1}.

We have first performed extensive search for stable geometries of silicene on Ag (111) surface. It is recognized that the 3 $\times$ 3 super-periodicity of silicene is commensurate to the 4 $\times$ 4 Ag(111) with the lateral Si spacing of 2.23 {\AA}, whereas the $\sqrt{7} \times \sqrt{7}$ periodicity of silicene is to the $\sqrt{13}\times \sqrt{13}$ and the $2\sqrt{3} \times 2\sqrt{3}$ Ag(111) with the lateral Si spacing of 2.28 {\AA} and 2.19 {\AA}, respectively. We have found two distinctive 
structures each for the 4$\times$4, the $\sqrt{13}\times\sqrt{13}$ and the $2\sqrt{3}\times2\sqrt{3}$ periodicities. The structural parameters of obtained stable (S) and metastable (MS) structures are shown in Table \ref{structure} along with the corresponding adsorption energy $E_{\rm ad}$, which is defined by $E_{\rm ad} = (E_{\rm Ag} + N_{\rm Si} \mu_{\rm Si} - E_{\rm Si/Ag}) / N_{\rm Si}$. Here $E_{\rm Si/Ag}$ and $E_{\rm Ag}$ are the total energies of silicene/Ag(111) and clean Ag(111), respectively, and $N_{\rm Si}$ is the number of Si atoms in the silicene. As for the chemical potential $\mu_{\rm Si}$ of Si, we use the total energy of an isolated Si atom. Calculated adsorption energies for the S and MS structures for the $\sqrt{13} \times \sqrt{13}$ and the 4 $\times$ 4 surfaces are close to each other: $E_{\rm ad}$ is largest for the S structure of the $\sqrt{13} \times \sqrt{13}$, followed by those for other three structures with less than 10-meV decrease. It is of note that $E_{\rm ad}$ of the silicene is comparable with the cohesive energy of Si in diamond structure (6.02 eV), and larger than that of free-standing silicene (5.27 eV in our calculations). The adsorption energies for the $2 \sqrt{3} \times 2 \sqrt{3}$ surface is relatively smaller by 50 - 70 meV, reflecting the compression of the Si layer on this surface. The energetics obtained is consistent with that the $2 \sqrt{3} \times 2 \sqrt{3}$ surface is less frequently observed \cite{vogt,lin, jamagotchian, lalmi, chen}.

\begin{table}
\caption{
Calculated adsorption energy $E_{\rm ad}$ (eV/Si-atom) and structural parameters of the stable(S) and the metastable (MS) structures for silicene on Ag(111) surfaces with the periodicities of 4$\times$4, $\sqrt{13} \times \sqrt{13}$ and $2 \sqrt{3} \times 2 \sqrt{3}$. The $d_{\rm Si-Si}$ and $d_{\rm Ag-Si}$ are the average distances (\AA) between Si atoms and between bottom Si and Ag surfaces. The $\Delta z$ ( \AA ) shows the Si coordinates in the surface-normal direction relative to that of the bottom Si.
}
\label{structure}
\begin{tabular}{cccccccc}
  \hline
  \hline
Periodicity  & str  & $E_{\rm ad}$  & $d_{\rm Si-Si}$ & $d_{\rm Ag-Si}$ & $\Delta z$ \\
  \hline
  4$\times$4 & S  & 5.972   & 2.29  & 2.13  & ---, ---, 0.78    \\
             & MS & 5.968   & 2.29  & 2.11  & 0.57, 0.70, 0.90  \\
  $\sqrt{13} \times \sqrt{13}$
             & S  & 5.975   & 2.32  & 2.13  & 0.36, 0.59, 0.79 \\
             & MS & 5.971   & 2.31  & 2.13  & 0.19, 0.52, 0.90 \\
  $2 \sqrt{3} \times 2 \sqrt{3}$
             & S  & 5.922   & 2.27  & 2.16  & ---, 0.36, 1.19 \\
             & MS & 5.906   & 2.27  & 2.21  & ---, ---, 1.06 \\
  \hline
  \hline
\end{tabular}
\end{table}

One of the characteristics common to all the S and MS structures is the buckling of silicene: The surface-normal coordinate $z$ of Si has 2 - 4 different values depending on the structures, as shown in Table \ref{structure}. The amount of the buckling is larger on average than that of free-standing silicene. The average of the Si-Si distance $d_{\rm Si-Si}$ is larger than that of free-standing silicene (2.24 {\AA} in our calculation). It is of note that the averaged distance between Si and Ag surface infers the formation of chemical bonds between silicene and Ag.

\begin{figure}
\begin{center}
\includegraphics[angle= 0, width=0.98\linewidth]{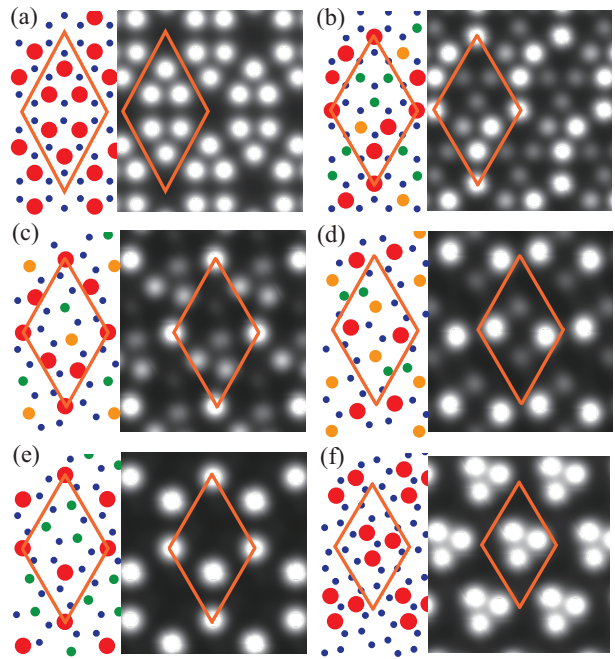}
\caption{
(color online) Calculated STM images (right parts) and lateral atomic positions of Si atoms (left parts) of stable structures for silicene/Ag(111) surfaces. The S (a) and the MS (b) structures for the 4 $\times$ 4 surface, the S (c) and the MS (d) structures for the $\sqrt{13} \times \sqrt{13}$, and the S (e) and the MS (f) structures for the $2\sqrt{3} \times 2\sqrt{3}$. The buckling of the silicene is represented by the size of each ball in the left parts: The largest red and smallest blue balls depict the top and the bottom Si atoms and other balls do the Si atoms between. The lateral unit cells are indicated by the orange lines.
}
\label{stm}
\end{center}
\end{figure}

The complex pattern of the buckling we find leads to a rich variety of STM images. Fig. \ref{stm} shows calculated STM images \cite{stmcal} of the S and MS structures for the silicene/Ag with 4 $\times$ 4, $\sqrt{13} \times \sqrt{13}$, and 2 $\sqrt{3} \times 2 \sqrt{3}$ periodicities. The calculated STM image for the S structure with each periodicity is in excellent agreement with the experiments \cite{vogt,lin,jamagotchian,feng}. Our S structure for the 4 $\times $ 4 surface seems essentially identical to the structures proposed in the previous works \cite{vogt,lin}. The structures with other periodicities have been proposed \cite{lin,feng} but the agreement with the experimental STM images is poor. On the contrary, the calculated STM images in the present work reproduce characteristic features in the experiments: The moderately bright region of the lower half of the lateral cell in the $\sqrt{13} \times \sqrt{13}$ structure [Fig. \ref{stm}(c)] and the super-hexagonal structure in the 2 $\sqrt{3} \times 2 \sqrt{3}$ structure [Fig. \ref{stm}(e)]. The calculated adsorption energy of each structure corroborates the identification proposed here. The MS structure with each periodicity exhibits the STM image different from that of the S structure [Figs. \ref{stm}(b), (d), (f)]. These images are not observed so far, but are likely to be observed in future since the difference in $E_{\rm ad}$ between the S and the MS structures is small.

\begin{figure}
\begin{center}
\includegraphics[angle= 0,width=0.9\linewidth]{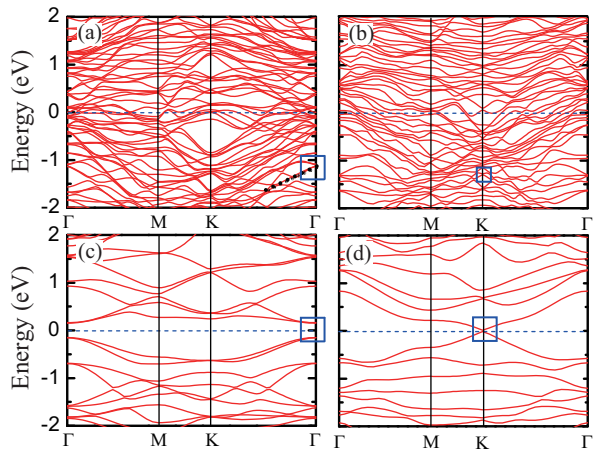}
\caption{
(color online) Calculated energy bands of the S structures for the 4 $\times$ 4 (a) and the $ \sqrt{13} \times \sqrt{13}$ (b) silicene/Ag(111) surfaces in the lateral BZ. The K point in the BZ of 1 $\times$ 1 silicene is folded on $\Gamma$ in the 4 $\times$ 4 and on K in the $ \sqrt{13} \times \sqrt{13}$ silicene/Ag(111) surfaces, respectively. $E_F$ is shown by dashed lines. Calculated energy bands of the isolated silicene layer peeled from the S structure of the 4 $\times$ 4 (c) and the $ \sqrt{13} \times \sqrt{13}$ (d) surfaces are also shown. States marked by squares have characters of $\pi$ and $\pi^*$, or the mixed $\pi$ and $\pi^*$ (see text). The linear band of mixed $\pi$ ($\pi^*$) state in (a) is indicated by the black dots.
}
\label{bands}
\end{center}
\end{figure}

\begin{figure}
\begin{center}
\includegraphics[angle= 0,width=0.9\linewidth]{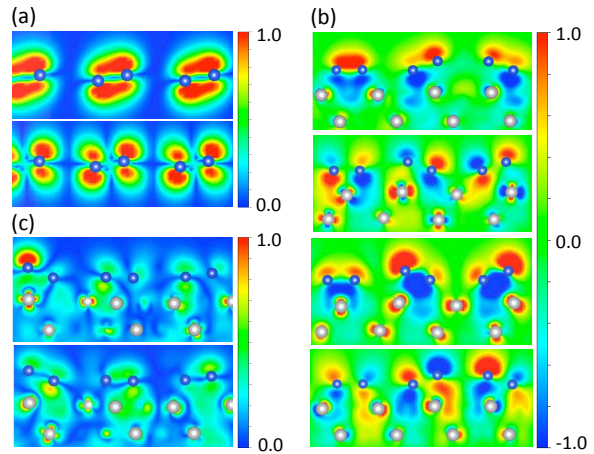}
\caption{(color online)
Contour plots of KS orbitals of the mixed $\pi$ ($\pi^{*}$ ) states in silicene/Ag(111) on the plane perpendicular to the silicene sheet. (a) KS orbitals (absolute value) of the $\pi$ (upper panel) and $\pi^{*}$ (lower panel) states for the freestanding silicene at K point in 1$\times$1 BZ. (b) KS orbitals of the four mixed $\pi$ ($\pi^{*}$) states marked in Fig. \ref{bands}(a) in the 4 $\times$ 4 silicene/Ag(111) at $\Gamma$ point. (c) KS orbitals (absolute value) of the two mixed $\pi$ ($\pi^{*}$) states marked in Fig. \ref{bands}(b) in the $\sqrt{13} \times \sqrt{13}$ silicene/Ag(111) at $K$ point. The blue and white balls depict Si and Ag atoms, respectively. The contour values are normalized to the maximum.
}
\label{KS}
\end{center}
\end{figure}

We are now in a position to discuss the electronic structure. Figure \ref{bands} shows the energy bands of the S structures of the 4 $\times$ 4 and the $\sqrt{13} \times \sqrt{13}$ silicene/Ag(111) surfaces. We also show the energy bands of the isolated 4 $\times$ 4 and the $\sqrt{13} \times \sqrt{13}$ silicene layers which are peeled from the Ag substrate in the S structures. In the low-buckled free-standing silicene, the energy bands clearly show Dirac cone and the corresponding Kohn-Sham (KS) orbitals at K point have characters of the bonding $\pi$ and the anti-bonding $\pi$ (labeled as $\pi^*$), as is shown in Fig. \ref{KS}(a).  In the isolated silicene peeled from the Ag substrate, the $\pi$ and $\pi^*$ states still exist near $E_F$ [Figs. \ref{bands}(c) and (d)]. They are almost degenerate at $E_F$ for the $\sqrt{13} \times \sqrt{13}$ structure, whereas they split for the 4 $\times$ 4 structure, opening a gap of 0.3 eV. It is recognized that the high buckling opens a gap due to incorporation of $sp^3$ hybridization \cite{cahangirov, fleurence}. The energy bands of the isolated 4 $\times$ 4 silicene is the case (see $\Delta z$ in Table \ref{structure}). In the isolated $\sqrt{13} \times \sqrt{13}$ silicene, the gap is very small presumably due to uneven arrangement of Si atoms with small $\Delta z$.

In real silicene/Ag(111), the $\pi$ and $\pi^*$ states disappear near $E_F$. Detailed analyses of the Kohn-Sham orbitals clarify that the states which have the components of $\pi$ or $\pi^*$ orbitals exist in the valence bands. They are at -1.11 eV (-1.28 eV) below $E_F$ for the 4 $\times$ 4 (the $\sqrt{13} \times \sqrt{13}$) structure [Figs. \ref{bands} (a) and (b)]. Our calculation shows that the amount of the electron transfer from Ag surface to silicene is less than 0.43 (0.36) electron per unit cell for the 4$\times$4 ($\sqrt{13}\times\sqrt{13}$) surface. This electron transfer causes the shift of $E_F$ to the higher energy bands of silicene by about a few tenths of eV, which is much smaller than the calculated shift of possible Dirac points.

Figures \ref{KS}(b) and (c) show KS orbitals of the above states. It is found that those $\pi$ and $\pi^*$ states strongly mixed with Ag orbitals at the top two layers [mixed $\pi$ and $\pi^*$ states]. This mixing leads to the formation of Si-Ag bonds in the S structures, rendering the states deep in the valence bands. The anti-bonding counter parts indeed exist around 1 eV above $E_F$ with the bonding-antibonding splitting being more than 2 eV. Calculated total energy gain, - ($ E_{\rm Si/Ag} - E_{\rm Ag} - E_{\rm silicene}$), is about 0.70 eV/Si-atom in both the 4 $\times$ 4 and $\sqrt{13} \times \sqrt{13}$ surfaces. Dirac states are substantially modified and shifted from $E_F$ due to the mixing of the Ag substrate. We have confirmed that the absence of Dirac electrons found here is common to all the (meta)stable structures of silicene on Ag(111).

\begin{figure}
\begin{center}
\includegraphics[angle= 0,width=0.9\linewidth]{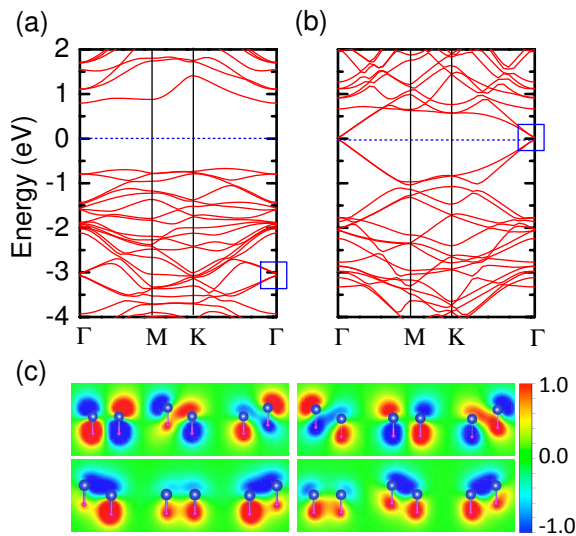}
\caption{
(color online) Calculated energy bands of hydrogen attached silicene 3 $\times$ 3/H$_{18}$ (a) and of flattened silicene 3 $\times$ 3 without H (b). The Kohn-Sham orbitals for the four mixed $\pi$ and $\pi^*$ states marked by the squares at $\Gamma$ in (a) are shown on the plane perpendicular to the silicene sheet in (c). The Dirac point comes up at $E_F$ in (b) depicted as a square. The blue and pink balls depict Si and H atoms, respectively.
}
\label{bandKS_SiH}
\end{center}
\end{figure}

In order to clarify the reason of the absence of Dirac electrons, we peel the silicene layer from the 4 $\times$ 4 Ag substrate, as in Fig. \ref{bands} (c), and then put hydrogen atoms  below the silicene with a fixed Si-H distance of 1.50 {\AA} as substitutes for the Ag atoms. The energy bands and the KS orbitals of the resultant 3 $\times $ 3 silicene/H$_{18}$ are shown in Figs. \ref{bandKS_SiH} (a) and (c). We find that the mixed $\pi$ ($\pi^*$) states appear around -3.1 eV at $\Gamma$ but the degeneracy is lifted by 57 meV similar to that in Fig. \ref{bands} (a). The bonding between Si and H atoms renders the mixed $\pi$ ($\pi^*$) states far from $E_F$. Energy bands of the silicene peeled from $\sqrt{13} \times \sqrt{13}$ Ag substrate with the H atoms attached show the essentially identical feature. We next make the H-attached 3 $\times$ 3 silicene/H$_{18}$ flat and examine the band structures. The gap between the lifted mixed $\pi$ ($\pi^*$) states located at around -3.1 eV becomes 25 meV in this case, recovering almost the linear dispersion. This small gap of 25 meV is due to the lateral distortion of Si hexagons in silicene. We finally remove H atoms from the above flat silicene 3 $\times $ 3 silicene/H$_{18}$. It is then found that the mixed $\pi$ and $\pi^*$ states lose the character of hydrogen orbitals and shift upward at $E_F$ [Fig.\ref{bandKS_SiH}(b)], with Dirac electrons recovered.

We now recognize that two factors are important in disappearance of Dirac electrons in silicene: the buckling of the silicene layer and the mixing with the substrate orbitals. Our calculations show that the buckling induces the mixing of the $\pi$ and the $\sigma$ orbitals and usually breaks the hexagonal symmetry, thus breaking the linear dispersion and the lifting of the degeneracy. When the silicene stands alone, the $\pi$ and $\pi^*$ states, whichever is mixed or not with the $\sigma$, stay near $E_F$. Then when the silicene is placed on the substrate, the mixing with the substrate orbitals induces the mixed states which are shifted downwards, being located deep in the valence bands. Interestingly, the mixing between $\pi$ ($\pi^*$) and the substrate orbitals have another effect: In silicene 4 $\times$ 4 /Ag (111), the mixing between Si and Ag orbitals reduces the mixing of the $\pi$ and the $\sigma$ orbitals in silicene and recovers the linear dispersion presumably due to the partial recovery of the hexagonal symmetry [Fig. \ref{stm} (a) and Fig. \ref{bands} (a)].

\begin{figure}
\begin{center}
\includegraphics[angle=0.0,width=0.95\linewidth]{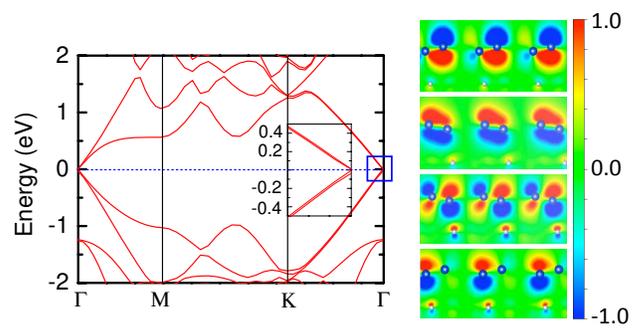}
\caption{
(color online) Calculated energy bands and contour plots of the mixed $\pi$ and $\pi^*$ states at $\Gamma$ point of silicene on h-BN. Extended energy bands near $E_F = 0$ are shown in the inset. The mixed $\pi$ and $\pi^*$ marked by the square are shown on the plane perpendicular to the silicene layer. The blue balls depict Si atoms.
}
\label{BN}
\end{center}
\end{figure}

\begin{figure}
\begin{center}
\includegraphics[angle=0.0,width=1\linewidth]{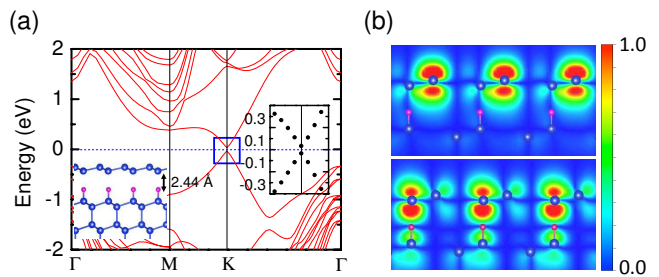}
\caption{
(color online) Calculated energy bands (a) of hydrogen-inserted Si (111) surface [inset of (a)] and contour plots of the mixed $\pi$ and $\pi^*$ states (b). The blue and pink balls in the inset depict Si and H atoms, respectively. Extended energy bands near $E_F = 0$ are shown in another inset and the distributions of the $\pi$ ($\pi^*$) states at K point marked by the square are shown on the plane perpendicular to the surface. 
}
\label{smartcut}
\end{center}
\end{figure}

Finally, we propose other two systems in which the character of Dirac electrons of silicene is satisfactorily preserved: i.e., the h-BN substrate and the hydrogen-processed Si. Figure \ref{BN} shows calculated energy bands and KS orbitals of silicene on h-BN sheet. In our calculation, $\sqrt{3} \times \sqrt{3}$ periodicity of silicene is commensurate with $\sqrt{7} \times \sqrt{7}$ periodicity of h-BN with the lattice mismatch of 0.64 \%, and the binding energy between silicene and h-BN is 56 meV/Si-atom with the optimized inter-layer distance of 3.23 {\AA}. The mixed $\pi$ and $\pi^*$ states appear near $E_F$ with the tiny gap of 10 meV, exhibiting the linear dispersion peculiar to Dirac electrons.

Figure \ref{smartcut} shows calculated energy bands and KS orbitals of Si (111) surface with the hydrogen monolayer intercalated in the subsurface. The H insertion in the subsurface is used to cut Si crystal \cite{smart-cut} and the single-layer H insertion may be possible in future. We have indeed found that such intercalated structure is stable with the separation between the top Si layer and the H layer of 2.44 {\AA}. The binding energy of the top Si layer to the H-covered Si (111) surface is 140 meV/Si-atom. The top Si layer keeps the hexagonal structure with the difference in the surface-normal coordinates $\Delta$z = 0.48 {\AA} of the Si atoms. Calculated energy bands and the KS orbitals clearly show the existence of Dirac electrons at $E_F$.

In conclusion, we have performed first-principles total-energy electronic-structure calculations for silicene on Ag (111) surfaces. We have found that buckling of the silicene layer and the mixing of $\pi$ and $\pi^*$ orbitals of silicene with the Ag orbitals are prominent. This makes the Dirac electrons of the hypothetical free-standing silicene disappear near the Fermi level. We have instead proposed that either BN substrate or the hydrogen intercalation in Si (111) subsurface provide a new stage to preserve Dirac electrons in real silicene.

\begin{acknowledgments}
This work was supported by the Grants-in-Aid under the contract number 22104005 and by CMSI, conducted by MEXT, Japan . Computations were performed mainly at Supercomputer Center in ISSP, UT. ZX acknowledges the support of NSFC (No. 11204259).
\end{acknowledgments}


\begin{thebibliography}{}
\bibitem{foot1}
Calculational parameters adopted here provides the lattice constant for the face-centered-cubic Ag of 4.03 {\AA}, 1.47 \% shorter than the experimental value. The maximum spacing between the adjacent $k$ points in Brillouine integral is less than 0.05 {\AA}$^{-1}$ in all the calculations.
\end{thebibliography}
\end{document}